\documentclass[letterpaper,twocolumn,prl,aps,superscriptaddress,floatfix]{revtex4-2}
\usepackage{mathrsfs}
\usepackage{mathptmx}

\usepackage[latin9]{inputenc}
\usepackage{amsmath}
\usepackage{graphicx}
\usepackage{esint}
\usepackage{diagbox}
\usepackage{multirow}
\usepackage{algpseudocode}
\usepackage{algorithm}
\usepackage{ulem}
\usepackage{lineno}
\modulolinenumbers[1]

\makeatletter
\pdfpageheight\paperheight
\pdfpagewidth\paperwidth

\@ifundefined{textcolor}{}
{%
 \definecolor{BLACK}{gray}{0}
 \definecolor{WHITE}{gray}{1}
 \definecolor{RED}{rgb}{0,0,0}
 \definecolor{GREEN}{rgb}{0,1,0}
 \definecolor{BLUE}{rgb}{0,0,1}
 \definecolor{CYAN}{cmyk}{1,0,0,0}
 \definecolor{MAGENTA}{cmyk}{0,1,0,0}
 \definecolor{YELLOW}{cmyk}{0,0,1,0}
}
\usepackage{xcolor}
\usepackage{soul}

\usepackage{soul}

\newcommand{\bra}[1]{\ensuremath{\left\langle#1\right|}}
\newcommand{\ket}[1]{\ensuremath{\left|#1\right\rangle}}
\definecolor{blue}{rgb}{0,0,1}
\definecolor{red}{rgb}{1,0,0}
\definecolor{green}{rgb}{0,1,0}

\usepackage[unicode=true,pdfusetitle,
 bookmarks=true,bookmarksnumbered=false,bookmarksopen=false,
 breaklinks=false,pdfborder={0 0 0},pdfborderstyle={},backref=false,colorlinks=true]
 {hyperref}
\hypersetup{
 colorlinks,linkcolor=red,citecolor=blue}

\makeatother

\begin{document}

\title{Universal two-qubit control across dilution refrigerators via a 10-km coherent optical link}

\author{Chuanlong~Ma}
\thanks{These three authors contributed equally to this work.}
\affiliation{Center for Quantum Information, Institute for Interdisciplinary Information Sciences, Tsinghua University, Beijing 100084, China}

\author{Jiajun Chen}
\thanks{These three authors contributed equally to this work.}
\affiliation{Center for Quantum Information, Institute for Interdisciplinary Information Sciences, Tsinghua University, Beijing 100084, China}

\author{Jia-Qi~Wang}
\thanks{These three authors contributed equally to this work.}
\affiliation{Laboratory of Quantum Information, University of Science and Technology of China, Hefei, Anhui 230026, China}

\author{Zheng-Xu~Zhu}
\affiliation{Laboratory of Quantum Information, University of Science and Technology of China, Hefei, Anhui 230026, China}

\author{Jia-Hua~Zou}
\affiliation{Laboratory of Quantum Information, University of Science and Technology of China, Hefei, Anhui 230026, China}

\author{Linze~Li}
\affiliation{School of Information Science and Technology, ShanghaiTech University, Shanghai, China}

\author{Dingran~Gu}
\affiliation{School of Information Science and Technology, ShanghaiTech University, Shanghai, China}

\author{Yuan-Hao~Yang}
\affiliation{Laboratory of Quantum Information, University of Science and Technology of China, Hefei, Anhui 230026, China}

\author{Lintao~Xiao}
\affiliation{Center for Quantum Information, Institute for Interdisciplinary Information Sciences, Tsinghua University, Beijing 100084, China}

\author{Xiaoxuan~Pan}
\affiliation{Center for Quantum Information, Institute for Interdisciplinary Information Sciences, Tsinghua University, Beijing 100084, China}

\author{Xin-Biao~Xu}
\affiliation{Laboratory of Quantum Information, University of Science and Technology of China, Hefei, Anhui 230026, China}
\affiliation{Hefei National Laboratory, Hefei 230088, China}

\author{Baile~Chen}
\affiliation{School of Information Science and Technology, ShanghaiTech University, Shanghai, China}

\author{Haifeng~Yu}
\affiliation{Beijing Academy of Quantum Information Sciences, Beijing 100193, China}
\affiliation{Hefei National Laboratory, Hefei 230088, China}

\author{Weiting~Wang}
\thanks{E-mail: wangwt2020@tsinghua.edu.cn}
\affiliation{Center for Quantum Information, Institute for Interdisciplinary Information Sciences, Tsinghua University, Beijing 100084, China}

\author{Chang-Ling~Zou}
\thanks{E-mail: clzou321@ustc.edu.cn}
\affiliation{Laboratory of Quantum Information, University of Science and Technology of China, Hefei, Anhui 230026, China}
\affiliation{Hefei National Laboratory, Hefei 230088, China}

\author{Luyan~Sun}
\thanks{E-mail: luyansun@tsinghua.edu.cn}
\affiliation{Center for Quantum Information, Institute for Interdisciplinary Information Sciences, Tsinghua University, Beijing 100084, China}
\affiliation{Hefei National Laboratory, Hefei 230088, China}

\date{\today}

\begin{abstract}
\textbf{
{Distributed quantum networks are important for scalable quantum processors and for potential applications in communication, distributed computation, and sensing, in which direct classical links for adaptive control are essential to harness the shared entanglement. However, controlling superconducting quantum processors from a separate refrigerator over long distances remains elusive. Here, we introduce a coherent optical link (COL) that delivers microwave waveforms and their phase references between dilution refrigerators. Combining a laser, a microwave-to-optical transducer and an optical-to-microwave photodiode, the COL transmits microwave control signals over 10~km of fiber, while optical references derived from the same laser compensate for the frequency mismatch between devices. Through the COL, we deliver a universal gate set to a transmon qubit and a cavity qubit in another refrigerator and entangle them, and achieve a Bell-state fidelity of 80.6\%. Quantum control delivered through the COL thus provides a classical backbone for modular quantum processors.}
}
\end{abstract}

\maketitle
\vskip 0.5cm
\noindent \textbf{\large{}Introduction}{\large\par} 
\noindent

Superconducting quantum processors are outgrowing the dilution refrigerators that house them. Every superconducting qubit requires its own control and readout lines, and their wiring, thermal loading, and volume limit the number of qubits that one cryostat can host~\cite{Krinner2019Engineeringcryogenicsetups,Bravyi2022}. Larger quantum systems will therefore be assembled from many cryogenic modules that operate as one, a distributed architecture that also underlies long-distance quantum communication, networks of entangled sensors, and quantum data centers~\cite{Jiang2007,Awschalom2021,Main2025,Guo2020,Zhang2021DQS,Liu2022}. Such modules need two kinds of links. Quantum links share entanglement between them, and classical links coordinate them by carrying the drives that set their gates and the feed-forward signals that act on mid-circuit measurements~\cite{Corcoles2021,CarreraVazquez2024,Chang2025}. Here, we focus on a coherent optical link (COL) for classical link, which converts a microwave control waveform into light, transmits it through fiber, and regenerates it in the receiving module with its phase preserved, so that the receiver obtains both the drive and its phase reference.

Delivering control as coherent light offers three advantages. First, optical fiber has low loss and low thermal conductance and enters a cryostat without the cascade of attenuators that every coaxial drive line requires, which reduces the power, size, and cost of control channels~\cite{Krinner2019Engineeringcryogenicsetups,Lecocq2021Controlreadoutsuperconducting,pan2025alloptical,Ma2025}. Second, the waveform arrives ready to drive the qubit. The receiving module needs no synthesizer or local oscillator, avoids the latency of decoding or digital-to-analog conversion in between. This direct path supports the low-latency classical communication required by teleported gates, dynamic circuits, and the adaptive coordination of several refrigerators~\cite{Gottesman1999,Chou2018,Corcoles2021,CarreraVazquez2024}. {Third, a COL is built from the same elements as a quantum link: a microwave-to-optical transducer, fiber, and photodetection. As transducers approach unit efficiency and less than one added noise photon~\cite{Lauk2020,Han2021Microwaveopticalquantum,Mirhosseini2020Superconductingqubitoptical,Higginbotham2018,Brubaker2022,Sahu2022Quantumenabledoperation,Sahu2023}, the same channel can carry heralded entanglement and, eventually, quantum states directly between refrigerators~\cite{Krastanov2021,Zou2025}. 
}

Optical delivery of control signals has advanced rapidly. Superconducting qubits have been driven and read out through optical fiber within a single refrigerator, with light modulated at room temperature and detected by cryogenic photodiodes~\cite{Lecocq2021Controlreadoutsuperconducting,Youssefi2021cryogenicelectrooptic,Xu2025Manipulationstransmonqubit,pan2025alloptical,Ma2025}, reaching single-qubit gate fidelities above 99.9\%~\cite{Li2026}, and a qubit has been driven through a microwave-to-optical transducer in the same refrigerator~\cite{Warner2025Coherentcontrolsuperconducting}. Links between separate refrigerators, however, remain rare. Cryogenic microwave links have connected refrigerators up to tens of meters apart~\cite{Magnard2020MicrowaveQuantumLink,Storz2023LoopholefreeBell,Kulikov2026}, and digital links connect processors in real time but carry bits, so every receiver needs its own synthesizer and an oscillator locked to the sender~\cite{CarreraVazquez2024,Chang2025,PeresScudo2001,Bartlett2007}. Recently, Zhou \textit{et al.} transferred coherent microwave pulses between superconducting resonators in two dilution refrigerators over 1\,km of fiber~\cite{Zhou2026}. They matched the frequencies of two electro-optic transducers by design and electrical tuning, derived the second pump from the first laser, and locked the microwave sources to a common reference. However, no qubit was driven through that link, and to our knowledge, a universal gate set, the basis of general-purpose distributed computation, has not been delivered from one refrigerator to another over kilometer-scale distances.

\begin{figure*}[ht]
\includegraphics{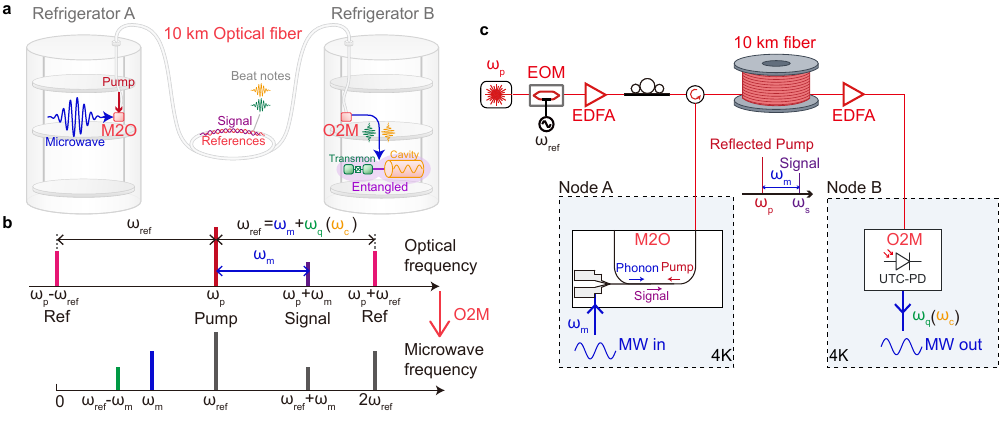}
\caption{\textbf{10-km coherent optical link.} \textbf{a,} Schematic of a $10~\mathrm{km}$ coherent optical link (COL) connecting two dilution refrigerators (A and B) for qubit and cavity control. A  M2O transducer in Refrigerator A converts microwave control signals into optical signals, which propagate together with the reference light through a $10~\mathrm{km}$ fiber to Refrigerator B. The beating of the signal and reference on a photodiode realizes the optical-to-microwave (O2M) conversion, generating microwave pulses that control a transmon qubit and its coupled storage cavity for entanglement. \textbf{b,} Frequency-conversion scheme. Upper panel: optical frequency components. A laser pump at $\omega_{\mathrm{p}}$ drives the M2O transducer to convert a microwave input at $\omega_{\mathrm{m}}$ into an optical signal at $\omega_{\mathrm{p}}+\omega_{\mathrm{m}}$. Intensity-modulation of the pump generates sidebands at $\omega_{\mathrm{p}}\pm\omega_{\mathrm{ref}}$, with the upper sideband serving as the optical reference. The modulation frequency is chosen as $\omega_{\mathrm{ref}}=\omega_{\mathrm{m}}+\omega_{\mathrm{q(c)}}$ for qubit (cavity) control. Lower panel: microwave beat notes generated by the O2M converter. Beating the signal with the reference produces the desired drive at $\omega_{\mathrm{ref}}-\omega_{\mathrm{m}}=\omega_{\mathrm{q(c)}}$, resonant with the qubit $\ket{g}\leftrightarrow\ket{e}$ (cavity $\ket{0}\leftrightarrow\ket{1}$) transition. Additional beat notes at $\omega_{\mathrm{m}}$, $\omega_{\mathrm{ref}}$, $\omega_{\mathrm{ref}}+\omega_{\mathrm{m}}$,and $2\omega_{\mathrm{ref}}$ result from beating of the signal with the reflected pump, the reference with the reflected pump, and the two references with each other, respectively. \textbf{c,} Experimental setup for the COL. EOM: electro-optic modulator; EDFA: erbium-doped fiber amplifier; MW: microwave; UTC-PD: uni-traveling-carrier photodiode.}
\label{fig:fig1}
\end{figure*} 

Here, we realize a COL between two dilution refrigerators connected by 10\,km of optical fiber 
and deliver through it a universal gate set for two qubits. A Brillouin microwave-to-optical (M2O) transducer~\cite{Wang2025,pan2025alloptical,Yang2025} in one refrigerator writes the control waveform, generated at room temperature in place of a cryogenic source, onto an optical signal. Optical references derived from the same pump laser travel with the signal to uni-traveling-carrier photodiodes (UTC-PDs) in the second refrigerator and set the frequency, phase, and amplitude of each drive, so that one transducer addresses targets at different frequencies. Through the COL, we perform single-qubit gates on a transmon, turn a storage cavity into a second qubit via Kerr blockade, and apply conditional gates between the two. With this gate set, delivered entirely from the first refrigerator, we prepare a Bell state of the two qubits, completing the chain from a waveform programmed in one refrigerator to an entangled state characterized in the other.

\smallskip{}
\noindent \textbf{\large{}Results}{\large\par}

\smallskip{}
\noindent\textbf{Coherent optical link between refrigerators}

\begin{figure*}[ht]
\includegraphics{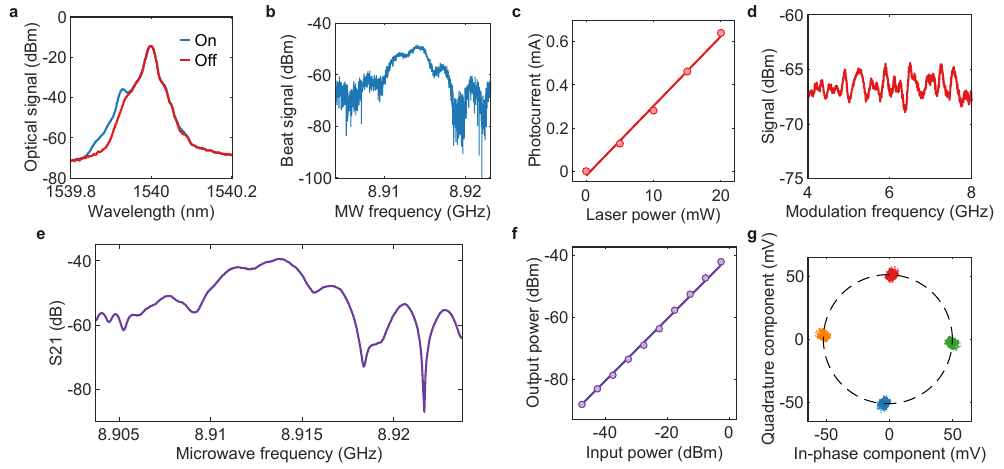}
\caption{\textbf{Characterization of the 10-km COL.}  
\textbf{a,} Optical spectra reflected from the M2O transducer with the microwave drive on and off. The converted signal lies 8.9~GHz above ({0.07\,nm} lower than) the pump. \textbf{b}, Conversion band of the M2O transducer, measured from the beat between the signal and the reflected pump on a room-temperature photodetector.  A Gaussian fit shows a full width at half maximum (FWHM) of 3.7~MHz. \textbf{c,} Photocurrent of the UTC-PD at zero bias versus incident optical power. Line, linear fit giving a responsivity of 0.0323~A\,W$^{-1}$. \textbf{d}, Microwave response of the UTC-PD to intensity-modulated light. \textbf{e,} Transmission $S_{21}$ of the full 10-km COL, from the input of the M2O transducer to the output of the O2M UTC-PD. A Gaussian fit shows a FWHM of 3.6~MHz. \textbf{f,} Output versus input microwave power at 8.9137~GHz. Line, fit with unit slope.
\textbf{g,} Demodulated output for four input phases separated by $\pi/2$. In \textbf{b, e, f,} and \textbf{g}, the reflected pump serves as the reference.}

\label{fig:fig2}
\end{figure*}

\begin{figure*}[ht]
\includegraphics{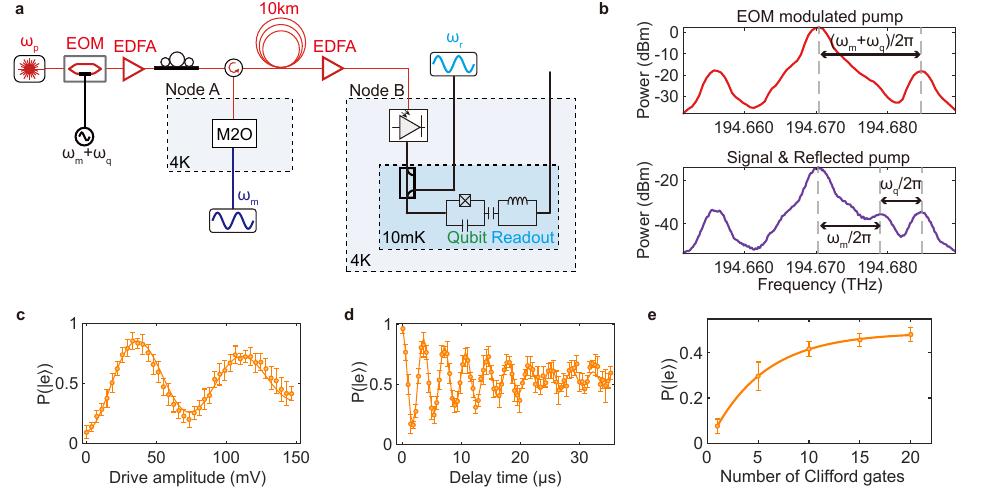}
\caption{\textbf{Remote control of a superconducting qubit through the 10-km COL.} 
\textbf{a,} Setup. An EOM sideband at $\omega_\mathrm{p}+\omega_\mathrm{m}+\omega_\mathrm{q}$ serves as the reference for the qubit drive. The qubit is read out through a local microwave line. \textbf{b,} Optical spectra of the EOM-modulated pump (top) and of the light reflected from the transducer (bottom), which contains the reflected pump, the signal, and the reference. \textbf{c}, Rabi oscillation versus the amplitude of a 4.8-$\mu$s Gaussian pulse. \textbf{d,} Ramsey fringes measured with two 1-$\mu$s Gaussian $\pi/2$ pulses. \textbf{e,} Single-qubit randomized benchmarking with 1-$\mu$s Gaussian pulses, giving an average fidelity of $(95.36\pm0.13)\%$ per elementary gate. Lines are fits, points and error bars are the mean and standard deviation of five (\textbf{c}, \textbf{d}) or 50 (\textbf{e}) repetitions.}

\label{fig:fig3}
\end{figure*}

\begin{figure*}[ht]
\includegraphics{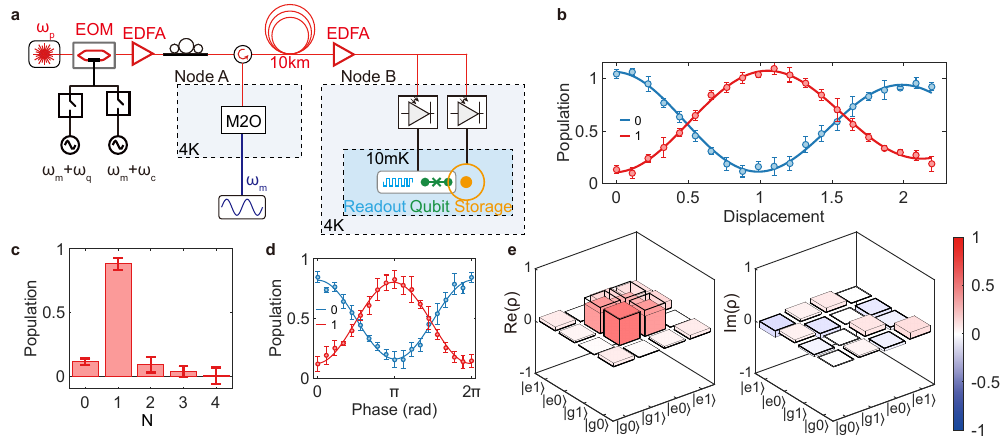}
\caption{\textbf{Remote control of a photonic qubit and its entanglement with the transmon.} \textbf{a,} Setup. Two switched EOM sidebands serve as references for the qubit and the cavity, which are driven by separate UTC-PDs. \textbf{b,} Populations of $|0\rangle$ and $|1\rangle$ versus the normalized amplitude of a 20-$\mu$s cavity pulse in the Kerr-blockade regime.
\textbf{c,} Photon-number distribution of the prepared state $|1\rangle$.
\textbf{d,} Ramsey interference of the cavity qubit. A SNAP gate between two cavity $\pi/2$ pulses imparts a phase $\theta$ to $|1\rangle$.
\textbf{e,} Real and imaginary parts of the reconstructed density matrix of the Bell state $(|g,1\rangle+|e,0\rangle)/\sqrt{2}$, prepared by a cavity $\pi/2$ pulse followed by a vacuum-selective transmon $\pi$ pulse. The state fidelity is 80.6\%. Wireframes show the ideal state. Preparation pulses are delivered through the COL. Photon-number readout and tomography use the local microwave lines, except in \textbf{c}, where the number-selective pulses are also delivered through the COL. Lines are fits; points and error bars are the mean and standard deviation of five repetitions.}

\label{fig:fig4}
\end{figure*} 

\noindent {Figure~\ref{fig:fig1}a sketches the system for delivering microwave control signals to a superconducting qubit-cavity system through a COL connecting two dilution refrigerators. Refrigerator~A (node~A) houses an M2O transducer~\cite{Wang2025,pan2025alloptical,Yang2025}  operated at approximately $4.3~\mathrm{K}$. Refrigerator~B (node~B) houses the  UTC-PDs~\cite{RN22,Li10475411,Chen2026Cryogenic,pan2025alloptical,Ma2025} for optical-to-microwave (O2M) conversion on its 4~K stage, which is operated at $3.4~\mathrm{K}$. The two nodes are connected by $10.1~\mathrm{km}$ of single-mode fiber. 
In refrigerator~A, the optically pumped M2O transducer converts microwave control pulses generated at room temperature into optical signals, which propagate through the fiber link together with the  optical reference. At refrigerator~B, beating between the signal and reference light in the UTC-PDs generates microwave pulses that drive a transmon qubit and its dispersively coupled storage cavity through cryogenic microwave lines, thereby enabling universal control of the coupled qubit-cavity system, including entangled-state preparation. Both the qubit and the cavity are located on the mixing-chamber plate at $11~\mathrm{mK}$. The COL is used for control, while qubit readout is performed locally using a conventional microwave chain.}

Figure~\ref{fig:fig1}c illustrates the operating principle of the COL. A laser at frequency $\omega_\mathrm{p}$ provides the optical pump for the M2O transducer. An electro-optic modulator (EOM),  driven at $\omega_\mathrm{ref}$, generates first-order sidebands at $\omega_\mathrm{p}\pm\omega_\mathrm{ref}$, with the upper sideband serving as the optical reference [Fig.~1b]. At node~A, a microwave drive at $\omega_\mathrm{m}$ applied to the M2O transducer launches an acoustic wave that scatters the pump, through phase-matched backward Brillouin scattering, into an counterpropagating optical signal at $\omega_\mathrm{p}+\omega_\mathrm{m}$ [inset of Fig.~1c]. Approximately $4\times10^{-4}$ of the optical power in  the pump and its sidebands is reflected at the optical interfaces of the chip and leaves the device together with the signal. An optical circulator routes the signal, the reflected pump, and  sidebands into the same fiber towards node~B. The frequency components of the output light are illustrated in the upper panel of Fig.~\ref{fig:fig1}b.

At node~B, the UTC-PD acts as a square-law detector, and the signal beats with the reference to produce a microwave photocurrent at $\omega_\mathrm{ref}-\omega_\mathrm{m}$ . The frequency of the delivered drive is  set by the EOM, and tuning $\omega_\mathrm{ref}$ to $\omega_\mathrm{m}+\omega_\mathrm{q}$ or $\omega_\mathrm{m}+\omega_\mathrm{c}$ addresses the qubit or the cavity with the same transducer. The beat note reproduces the envelope of the drive applied at node~A and carries its phase variations with the sign reversed, because the reference lies above the signal in frequency.  Because the signal and the references originate from the same laser and share one optical path, laser phase noise and fiber phase fluctuations are common to both and largely cancel in the beat note.

With the  chosen optical reference present, the O2M output contains a set of beat notes [lower panel of Fig.~\ref{fig:fig1}b].  Beating between the signal with the  reference produces the target microwave drive at $\omega_q$ or $\omega_\mathrm{c}$ to control the qubit or cavity, respectively.  Beating among the signal, reflected pump, and upper and lower sidebands also generates additional tones at $\omega_\mathrm{m}$, $\omega_\mathrm{ref}$, $\omega_\mathrm{ref}+\omega_\mathrm{m}$,and $2\omega_\mathrm{ref}$.  These additional tones lie gigahertz away from the qubit and cavity transitions and drive no transitions. Their residual effect is an off-resonant AC Stark shift, which we estimate to be sub-kHz and is negligible in the frequency calibration.

The M2O transducer is a straight lithium niobate waveguide that guides both light and sound, terminated by an interdigital transducer (IDT)~\cite{Yang2023StimulatedBrillouininteraction,Wang2025,Yang2025}. It is side-coupled to a lensed fiber that is aligned \textit{in situ} with a cryogenic nanopositioner (attocube). The IDT converts the microwave drive at $\omega_\mathrm{m}$ into a guided acoustic wave through the piezoelectric effect. Through the photoelastic effect, this wave scatters pump photons at $\omega_\mathrm{p}$ into anti-Stokes photons at $\omega_\mathrm{p}+\omega_\mathrm{m}$ that propagate opposite to the pump, a process known as phase-matched backward Brillouin scattering~\cite{Brillouin2019,Sipe2016Hamiltoniantreatmentstimulated,Yang2024ProposalBrillouinmicrowave}. Phase matching requires the acoustic wavevector to equal the sum of the two optical wavevectors, $q\simeq2n_{\mathrm{eff}}\omega_\mathrm{p}/c$, which sets the operating frequency at $\omega_\mathrm{m}/2\pi=2n_{\mathrm{eff}}v_\mathrm{a}/\lambda\approx8.9~\mathrm{GHz}$ for a fixed pump frequency $\omega_\mathrm{p}$, where $v_\mathrm{a}$ is the acoustic velocity and $n_{\mathrm{eff}}$ the effective index of the guided modes. Each anti-Stokes photon is created by annihilating one phonon of the driven acoustic wave, so the optical signal inherits the amplitude and phase of the microwave drive, and the photon-number conversion efficiency grows linearly with the on-chip pump power.

Figure~\ref{fig:fig2}a shows the optical spectra  of the light returning from the M2O transducer, measured using an optical spectrum analyzer (OSA) with the microwave drive on and off. When the drive is on, the anti-Stokes signal appears $8.9~\mathrm{GHz}$ above the reflected pump. {After accounting for the fiber-to-waveguide coupling efficiency $\eta_\mathrm{c}=1.4\%$ for both pump injection and signal extraction, together with the IDT microwave-to-phonon conversion efficiency $\eta_{\mathrm{IDT}}\approx5\%$,  we obtain an on-chip conversion efficiency, normalized to the on-chip pump power, of $\eta_{\mathrm{M2O}}={16\%}/\mathrm{W}$.} With a launched on-chip pump power $P_\mathrm{p}=1.3~\mathrm{mW}$, the photon-number efficiency from the IDT input to the fiber is $\eta_{\mathrm{sys}}=\eta_\mathrm{c}\eta_{\mathrm{IDT}}\eta_{\mathrm{M2O}}P_\mathrm{p}=1.5\times10^{-7}$. {In a separate measurement of the same M2O device, we obtain a higher normalized efficiency of $47\%/W$ at a pump wavelength of 1560~nm, achieving an on-chip efficiency of 0.12\% with a pump power $P_\mathrm{p}=2.5~\mathrm{mW}$.} Sweeping the microwave drive frequency and detecting the beat note between the signal and the reflected pump with a high-speed photodetector operating at room temperature yields the conversion spectrum shown in Fig.~\ref{fig:fig2}b. A Gaussian fit gives a full width at half maximum of $3.7~\mathrm{MHz}$. It corresponds to a response time of about $0.1~\mu\mathrm{s}$, so microsecond control pulses pass without distortion, and the transducer sets the bandwidth of the whole COL.

At node~B, the O2M converter is an array of UTC-PDs~\cite{RN22,Li10475411,pan2025alloptical,Ma2025,Chen2026Cryogenic}. In a UTC-PD, only electrons traverse the depleted collector, which gives a fast response and a high saturation current even without bias. We therefore operate the photodiodes at zero bias, which requires no DC wiring and dissipates no bias power at 4~K. The photocurrent of the element used for qubit control grows linearly with the incident optical power up to 20~mW, with a DC responsivity $R_{\mathrm{DC}}=0.0323~\mathrm{A/W}$, as shown in Fig.~\ref{fig:fig2}c. Its microwave response to intensity-modulated light shows no roll-off between 4 and 8~GHz as shown in Fig.~\ref{fig:fig2}d, with a 3-dB bandwidth at zero bias exceeding 9~GHz. A reverse bias would extend the bandwidth and the saturation current further, but the present band already covers the qubit and cavity frequencies, $\omega_\mathrm{q}/2\pi=5.3873~\mathrm{GHz}$ and $\omega_\mathrm{c}/2\pi=5.5965~\mathrm{GHz}$, at which we place the references. The $\pm2~\mathrm{dB}$ ripple in Fig.~\ref{fig:fig2}d has periods exceeding $200\,\mathrm{MHz}$,  {which may arise from impedance mismatches in the microwave components and ports of the measurement chain.} Since the periods are much larger than the MHz bandwidth of the M2O transducer, this effect only rescales the drive amplitude at each carrier frequency.

Connecting the two nodes completes the COL as shown in Fig.~\ref{fig:fig1}c. We characterize the COL by directly measuring the beat note between the pump and the signal at $8.9~\mathrm{GHz}$, i.e., with all other reference tones turned off. As shown in Fig.~\ref{fig:fig2}e, the link response spectrum agrees with that of the M2O device, with a bandwidth of $3.6~\mathrm{MHz}$. The output power follows the input power with a unit slope over 40~dB, as shown in Fig.~\ref{fig:fig2}f. The COL provide a transmittance of the microwave output over the input port at distinct cryostats of {$-39.4\,\mathrm{dB}$}, which is about $45\,\mathrm{dB}$ higher than that reported for the 1-km link between electro-optic M2O transducers~\cite{Zhou2026}. Stepping the input phase by $\pi/2$ produces four IQ clusters separated by $\pi/2$ as shown in Fig.~\ref{fig:fig2}f,  with an rms phase fluctuation of $0.042~\mathrm{rad}$ for an integration time of $100~\mathrm{\mu s}$ per point. The delivered phase is robust because the signal and the reference share one fiber. A change $\delta L$ in the common optical path delays both beams equally and shifts the phase of the beat note by only $0.11~\mathrm{rad}$ for a path difference of one millimeter at $\omega_\mathrm{q}$. 

\smallskip{}
\noindent\textbf{Remote control of a transmon qubit.}

\noindent We next characterize remote transmon control using the COL, as shown in Fig.~\ref{fig:fig3}a.  An optical reference at $\omega_\mathrm{p}+\omega_\mathrm{m}+\omega_\mathrm{q}$ beats with the optical signal of the M2O transducer at $\omega_\mathrm{p}+\omega_\mathrm{m}$ upon photodetection,  generating a resonant drive at the qubit $\ket{g}\leftrightarrow\ket{e}$ transition frequency $\omega_\mathrm{q}/2\pi$. The input EOM-modulated pump (before EDFA) and the reflected output of the M2O device are measured with an OSA, as shown in Fig.~\ref{fig:fig3}b, resolving the reference at $\omega_\mathrm{p}+\omega_\mathrm{m}+\omega_\mathrm{q}$ and the converted signal at $\omega_\mathrm{p}+\omega_\mathrm{m}$. The apparent widths of these optical tones are limited by the spectral resolution of the analyzer. The UTC-PD output and the local microwave drive are combined using a directional coupler, allowing the transmon to be driven through either path. Qubit control and  readout share a common input port, with readout always performed through the local microwave chain. 

Figures~\ref{fig:fig3}c and \ref{fig:fig3}d  demonstrate remote control of the qubit through the 10-km COL via Rabi and Ramsey measurements, respectively. For the Rabi measurement, we vary the amplitude of the microwave input to the M2O transducer while keeping the Gaussian pulse duration fixed at $4.8~\mu\mathrm{s}$, with $\sigma=1.2~\mu\mathrm{s}$ and truncation at $\pm2\sigma$. The resulting oscillations in the excited-state population demonstrate control over the qubit rotation angle. {The damping is caused by quasi-static fluctuations in the effective drive amplitude arising from intensity fluctuations in the reflected optical reference.} The Ramsey experiment consists of two Gaussian $\pi/2$ pulses separated by a variable delay, each with a duration of $1~\mu\mathrm{s}$ and $\sigma=0.25~\mu\mathrm{s}$. The phase of the second pulse is advanced linearly with the delay to produce the observed Ramsey fringes. Fitting the fringe decay yields $T_2=16\pm1~\mu\mathrm{s}$.


We further evaluate the fidelity of qubit control through the COL using single-qubit Clifford randomized benchmarking, as shown in Fig.~\ref{fig:fig3}e. Each sequence contains $m$ random selected Clifford gates followed by a recovery gate. All elementary single-qubit gates constituting the Clifford gates sequence are implemented using Gaussian pulses delivered through the COL, each with a duration of $1~\mu\mathrm{s}$ and $\sigma=0.25~\mu\mathrm{s}$. Fitting the averaged excited-state population as a function of  the number of Clifford gates  yields an estimated average single-qubit gate fidelity of $(95.36\pm0.13)\%$. {The observed gate infidelity arises from relaxation, dephasing, and drive-amplitude fluctuations.}


\smallskip{}
\noindent\textbf{Remote control of the coupled qubit-cavity system}

\noindent
We extend the COL to control the storage cavity coupled to the transmon in Refrigerator B, as illustrated in Fig.~\ref{fig:fig4}a. The cavity-qubit system is described by  

\begin{equation}
\begin{aligned}
H=&\omega_\mathrm{c} \hat{a}_\mathrm{s}^\dagger\hat{a}_\mathrm{s}+\omega_\mathrm{q} \ket{e}\bra{e}\\&-\chi_\mathrm{sq} \ket{e}\bra{e} \hat{a}_\mathrm{s}^\dagger\hat{a}_\mathrm{s}-\frac{K_\mathrm{g}}{2}\hat{a}_\mathrm{s}^\dagger\hat{a}_\mathrm{s}^\dagger\hat{a}_\mathrm{s}\hat{a}_\mathrm{s},
\end{aligned}
\label{Eq1:systemH}
\end{equation}
where $\omega_\mathrm{c}$ ($\omega_\mathrm{q}$) is the frequency of the storage cavity mode $\hat{a}_\mathrm{s}$ (the transmon $\ket{g}\leftrightarrow\ket{e}$ transition), $\chi_\mathrm{sq}/{2\pi}=5.593\,\mathrm{MHz}$ is the cross-Kerr interaction between the storage cavity and the transmon, and $K_\text{g}/{2\pi}=124\,\mathrm{kHz}$ is the self-Kerr of the storage cavity. An additional optical reference at $\omega_\mathrm{p}+\omega_\mathrm{m}+\omega_\mathrm{c}$  beats with the optical signal generated by the M2O transducer to produce a microwave drive at the cavity frequency $\omega_\mathrm{c}$. Separate elements of the UTC-PDs array deliver the qubit and cavity drives after transmission through the 10-km fiber, allowing both degrees of freedom to be controlled from Refrigerator A. The qubit and cavity reference tones are gated by switches to ensure that they do not  overlap in time.

Figure~\ref{fig:fig4}b shows oscillations of the vacuum and one-photon populations as the amplitude of a  $20\,\mu\mathrm{s}$ square cavity pulse, which is normalized as the displacement amplitude, is varied. The cavity self-Kerr nonlinearity enables selective driving of the $\ket{0}\leftrightarrow\ket{1}$ transition in the weak-drive regime, defining a photonic qubit in the two lowest Fock states. Photon-number-selective transmon pulses applied through the local microwave line map these populations onto the transmon readout. {The damping of these Rabi oscillations is likewise caused by quasi-static fluctuations in the drive amplitude.} We then prepare the one-photon state and also deliver the photon-number-selective qubit pulses through the COL, as shown in Fig.~\ref{fig:fig4}c. The resulting distribution is dominated by the one-photon response, demonstrating remote cavity excitation and photon-number-selective transmon control. 

The dispersive coupling between the transmon and the photonic qubit enables the two-qubit gates, as the transmon transition frequency is shifted conditionally on the photonic qubit state. To demonstrate control of the two-qubit interaction, we apply a selective number-dependent arbitrary phase (SNAP) gate $S_1(\theta)=\exp(i\theta|1\rangle\langle1|)$ to induce a relative phase on the photonic qubit, as the transmon qubit is flipped twice conditioned on the photonic qubit state. By placing the SNAP gate between two Hadamard gates of the photonic qubit, we realize phase-controlled interference of the photonic qubit, with both the cavity and qubit control pulses delivered through the COL. Sweeping the SNAP phase $\theta$ produces the population fringes shown in Fig.~\ref{fig:fig4}d. Finally, we combine cavity control with a vacuum-selective transmon flip enabled by the dispersive coupling, with all state-preparation pulses delivered through the COL, to prepare the Bell state $|\Psi^+\rangle=(|g,1\rangle+|e,0\rangle)/\sqrt{2}$ at the receiving node. Figure~\ref{fig:fig4}e presents the reconstructed density matrix, showing a Bell-state fidelity $F=\langle\Psi^+|\rho_{\rm rec}|\Psi^+\rangle=80.6{\pm 13.1}\%$.  

\smallskip{}

\noindent \textbf{\large{}Discussion}{\large\par}

\noindent 
Our results establish a COL for delivering quantum-control waveforms between dilution refrigerators over $10\,\mathrm{km}$ of fiber. The co-propagation of signal and optical reference in the fiber enables a highly stable phase of the recovered microwave drive and allows flexible control of their frequencies to match the qubit transitions. The single-qubit operations and Bell-state preparation demonstrate that the COL is ready for cooperating cryogenic quantum processors over long distances. Although there is plenty of room to improve the system efficiency of the M2O, the present performance identifies coupling instability as an important source of control imperfections, and also reveals unconventional decoherence processes associated with the COL that warrant further investigation. The demonstrated COL validates the coherence properties of the M2O transducer and signal-reference over long-distance propagation, which also establishes the foundation for quantum links. With higher transduction efficiency, for example in cavity-enhanced Brillouin transducers~\cite{Yang2024ProposalBrillouinmicrowave}, the same fiber could carry quantum states and heralded entanglement between the refrigerators~\cite{Krastanov2021,Zou2025}, while the COL supplies the control that acts on them.

\inputencoding{utf8}
%

\vbox{}

\smallskip{}





\smallskip{}

\smallskip{}

\noindent \textbf{\large{}{}Acknowledgment}{\large\par}

\noindent This work was funded by the Quantum Science and Technology-National Science and Technology Major Project (Grant Nos.~2024ZD0301500 and 2021ZD0300200), the National Natural Science Foundation of China (Grant Nos.~92265210, 12404567, 123B2068, 12550006, 12474498, 92365301, 92565301, 12374361, and 12293053). We also acknowledge the support from the Fundamental Research Funds for the Central Universities and USTC Research Funds of the Double First-Class Initiative. The UTC photodiode chips were fabricated with support from the ShanghaiTech University Material and Device Lab (SMDL). The numerical calculations in this paper were performed on the supercomputing system in the Supercomputing Center of USTC, and this work was partially carried out at the USTC Center for Micro and Nanoscale Research and Fabrication.

\smallskip{}

\noindent \textbf{\large{}{}Author contributions}{\large\par}

\noindent {C.-L.Z. and L.S. conceived the experiments. C.M., J.C., and J.-Q.W. built the experimental setup, carried out the measurements, and analyzed the data with the assistance from Z.-X.Z., J.-H.Z., L.L., Y.-H.Y., L.X., X.P., X.-B.X., and W.W.. J.-Q.W., Z.-X.Z., Y.-H.Y., and J.-H.Z. designed and fabricated the M2O transducers. L.L. and B.C. designed and fabricated the UTC-PD arrays. H.Y. fabricated the transmon qubits. C.M., J.C., J.-Q.W., W.W., C.-L.Z. and L.S. wrote the manuscript with input from all authors. W.W., C.-L.Z., and L.S. supervised the project.}

\clearpage{}

\end{document}